\documentclass[twocolumn]{aastex62}
\usepackage{color,soul,appendix}

\PassOptionsToPackage{draft}{hyperref}
\usepackage{xspace}

\usepackage[xindy, toc, hyperfirst=false, nolist, nostyles, sanitize={name=false,description=false,symbol=false}]{glossaries}
\glsdisablehyper
\usepackage[hyperref,x11names, table]{}

\newglossaryentry{vrad}{name={radial velocity~}, text={radial velocity}, symbol={\ensuremath{v_\textrm{rad}}}, description={radial velocity}, sort=vrad}
\newglossaryentry{vrot}{name={stellar rotation~}, name={stellar rotation}, symbol={\ensuremath{v_\textrm{rot}}}, description={radial velocity}, sort=vrot}

\newcommand{\xray}{X-ray}

\newglossaryentry{angstrom}{name=\AA, description={unit of length $10^{-10}$\,m}, sort=angstrom}
\newglossaryentry{nir}{name=NIR,description={near infrared},first = {near infrared (NIR)}}
\newglossaryentry{psf}{name=PSF,description={point-spread function},first = {point-spread function (PSF)}}
\newglossaryentry{fwhm}{name=FWHM,description={Full Width Half Maximum},first = {FWHM}}
\newglossaryentry{rms}{name=RMS,description={Root Mean Square},first = {RMS}}
\newglossaryentry{signalnoise}{name=S/N,description={signal to noise}}
\newglossaryentry{uv}{name=UV,description={ultra violet},first = {ultra violet (UV)}}
\newglossaryentry{halpha}{name=\ensuremath{\textrm{H}\alpha}, description={First line of the Balmer series at 6563\,\AA}, sort=halpha}
\newglossaryentry{mgb}{name={Mg \textsc{i} b}, description={Triplet at 5167\,\AA, 5173\,\AA and 5184\,\AA}}
\newglossaryentry{sobolevapprox}{name={Sobolev approximation}, description={Lines are approximation with an infinitley thin interaction region \citep[e.g. no broadening][]{1960mes..book.....S}}, first={Sobolev approximation }}
\newglossaryentry{radeq}{name={radiative equilibrium}, description={The net flux of energy between matter and radiation field is zero}}
\newglossaryentry{nebularapprox}{name={nebular approximation}, description={Assumes that the plasma condition are controlled by a central radiation source. The radiation field decreases with the distance to the source by geometrical dilution. See \citet{1978stat.book.....M} for details}}
\newglossaryentry{modnebularapprox}{name={modified nebular approximation}, description={In contrast to \gls{nebularapprox} where only geometrical dilution is taken into account, the modified nebular approximation also takes dilution by other radiative processes into account }, first={modified nebular approximation}, parent=nebularapprox}
\newglossaryentry{thompsonscat}{name={Thomson scattering}, description={Scattering of photons on low energy electrons}}
\newglossaryentry{lte}{name={LTE}, description={Local Thermodynamic Equilibrium}, first={local thermodynamic equilibrium (LTE)}}
\newglossaryentry{lsr}{name={LSR}, description={Local Standard of Rest}, first={\textit{local standard of rest} (LSR)}}
\newglossaryentry{mc}{name={MC}, description={Monte Carlo}, first={\textit{Monte Carlo} (MC)}}
\newglossaryentry{wcs}{name={WCS}, description={world coordinate system}, first={world coordinate system (WCS)}}
\newglossaryentry{cmf}{name=CMF, text=CMF, first=Comoving Frame (CMF henceforth), description={Comoving Frame}}

\newglossaryentry{uvoir}{name=UVOIR, text=UVOIR, first=UV/optical/Near-IR (UVOIR), description={UV/optical/Near-IR}}

\newglossaryentry{sfit}{name=SFIT, text=\textsc{sfit}, description={spectral fitting program for hot stars \citep{2001A&A...376..497J}}, first={\textsc{sfit} \citep{2001A&A...376..497J}}}
\newglossaryentry{iraf}{name=IRAF, text=\textsc{iraf}, description={Image Reduction and Analysis Facility maintained by NOAO}, first={\textsc{iraf}\protect\footnote{IRAF: the Image Reduction and Analysis Facility is distributed by the National Optical Astronomy Observatory, which is operated by the Association of Universities for Research in Astronomy (AURA) under cooperative agreement with the National Science Foundation (NSF).}}}
\newglossaryentry{pyraf}{name=PyRAF, text=\textsc{PyRAF}, description={Python wrap of \gls{iraf} maintained by STSCI}, first=\textsc{PyRAF} \protect\footnote{PyRAF is a product of the Space Telescope Science Institute, which is operated by AURA for NASA.}}
\newglossaryentry{astropy}{name=ASTROPY, text=\textsc{astropy}, description=\textsc{astropy} framework, first = \textsc{astropy} \citep{2013A&A...558A..33A}}
\newglossaryentry{numpy}{name=NUMPY, text=\textsc{numpy}, description=\textsc{numpy} framework, first = \textsc{numpy} \citep{walt2011numpy}}
\newglossaryentry{scipy}{name=SCIPY, text=\textsc{scipy}, description=\textsc{scipy} framework, first = \textsc{scipy} \citep{Jones:2001fk}}
\newglossaryentry{matplotlib}{name=matplotlib, text=\textsc{matplotlib}, description=\textsc{matplotlib} framework, first = \textsc{matplotlib} \citep{hunter2007matplotlib}}
\newglossaryentry{pandas}{name=pandas, text=\textsc{pandas}, description=\textsc{pandas} framework, first = \textsc{pandas} \citep{mckinney2010data}}
\newglossaryentry{ipython}{name=ipython, text=\textsc{ipython}, description=\textsc{ipython} framework, first = \textsc{ipython} \citep{perez2007ipython}}
\newglossaryentry{jupyter}{name=jupyter, text=\textsc{jupyter}, description=\textsc{jupyter} framework, first = \textsc{jupyter} \citep{kluyver2016jupyter,perez2015project,ragan2014jupyter}}
\newglossaryentry{aplpy}{name=aplpy, text=\textsc{aplpy}, description=\textsc{aplpy} framework, first = \textsc{aplpy} \citep{2012ascl.soft08017R}}
\newglossaryentry{nltk}{name=nltk, text=\textsc{nltk}, description=\textsc{nltk} framework, first = Natural Language ToolKit \citep[\textsc{NLTK};][]{bird2009natural}}
\newglossaryentry{scikit-learn}{name=scikit-learn, text=\textsc{scikit-learn}, description=\textsc{scikit-learn} framework, first = \textsc{scikit-learn} \citep[][]{scikit-learn}}
\newglossaryentry{scikit-image}{name=scikit-image, text=\textsc{scikit-image}, description=\textsc{scikit-image} framework, first = \textsc{scikit-image} \citep[][]{scikit-image}}
\newglossaryentry{moog}{name=MOOG,text={\textsc{moog}}, description={spectral synthesis software \citep{1973ApJ...184..839S}}, first={\textsc{Moog} \citep{1973ApJ...184..839S}}}
\newglossaryentry{atlas9}{name=ATLAS9,description={grid of stellar atmospheres \citep{2004astro.ph..5087C}}, first={ATLAS9 \citep{2004astro.ph..5087C}}}
\newglossaryentry{vald}{name=VALD,description={Vienna Atomic Line Database \citep{2000BaltA...9..590K}}, first={Vienna Atomic Line Database \citep[VALD;][]{2000BaltA...9..590K}}}
\newglossaryentry{sextractor}{name=SExtractor, text=\textsc{SExtractor}, description={Source Extractor photometry program \citep{1996A&AS..117..393B}}, first={\textsc{SExtractor} \citep{1996A&AS..117..393B}}}

\newglossaryentry{swarp}{name=SWarp, text=\textsc{SWarp}, description={SWarp \citep{2002ASPC..281..228B}}, first={\textsc{SWarp} \citep{2002ASPC..281..228B}}}
\newglossaryentry{astrometry.net}{name=astrometry.net, text=\textsc{astrometry.net}, description={\textsc{astrometry.net} \citep{2010AJ....139.1782L}} first={\textsc{astrometry.net} \citep{2010AJ....139.1782L}}}

\newglossaryentry{astrodrizzle}{name=AstroDrizzle, text=\textsc{AstroDrizzle}, description={AstroDrizzle \citep{2012drzp.book.....G}}, first={\textsc{AstroDrizzle} \citep{2012drzp.book.....G}}}

\newglossaryentry{idl}{name=IDL,text={\textsc{idl}}, description={Interactive Data Language}}
\newglossaryentry{makee}{name=MAKEE,text=\textsc{makee}, description={MAuna Kea Echelle Extraction by Tom Barlow available}}
\newglossaryentry{minuit}{name=MINUIT,text={\textsc{minuit}}, description={collection of numerical optimization tools \citep{James:1975dr}}}
\newglossaryentry{migrad}{name=MIGRAD,text={\textsc{migrad}}, description={numerical gradient optimization tools - part of \gls{minuit}}}
\newglossaryentry{dolphot}{name=DOLPHOT, text=\textsc{dolphot}, description=photometry package for HST, first=\textsc{dolphot} \citep{2000PASP..112.1383D}}
\newglossaryentry{synphot}{name=synphot, text={\textsc{synphot}}, description={synthetic photometry package from STSCI}, first={\textsc{synphot}\protect\footnote{\textsc{synphot} is a product of the Space Telescope Science Institute, which is operated by AURA for NASA.}}}
\newglossaryentry{chianti}{name=CHIANTI, text=CHIANTI, description= CHIANTI Database 7.1, first =CHIANTI 7.1 \citep{1997A&AS..125..149D,2012ApJ...744...99L}}
\newglossaryentry{synpp}{name=SYNPP, text=SYN++, description= SYN++ software, first =SYN++ \citep{2011PASP..123..237T}}
\newglossaryentry{tardis}{name=TARDIS, text=\textsc{tardis}, description= TARDIS MC code, first = {\textsc{tardis} \citep{2014MNRAS.440..387K}}}
\newglossaryentry{hach12}{name=H12, text=H12, description= Hachinger et al 2012 paper on 94I and hidden He, first = {\citep[][henceforth referred to as H12]{hachinger94Imodel}}}

\newglossaryentry{artis}{name=ARTIS, text=\textsc{artis}, description= ARTIS MC code, first = \textsc{artis} \citep{2009MNRAS.398.1809K}}
\newglossaryentry{cmfgen}{name=CMFGEN, text=\textsc{cmfgen}, description=CMFGGEn radiative transfer code, first = \textsc{cmfgen} \citep{1998ApJ...496..407H}}
\newglossaryentry{sedona}{name=SEDONA, text=\textsc{sedona}, description= Sedona MC code, first = \textsc{sedona} \citep{2006ApJ...651..366K}}
\newglossaryentry{phoenix}{name=PHOENIX, text=\textsc{phoenix}, description= PHOENIX radiative transfer code, first = \textsc{phoenix} \citep{1997ApJ...490..803H}}
\newglossaryentry{mlmc}{name=MLMC, text=ML93, description= Mazzali Lucy Monte Carlo, first ={Mazzali \& Lucy (1993, ML93) code}}
\newglossaryentry{starkit}{name=STARKIT, text=\textsc{starkit}, description= TARDIS MC code, first = {\textsc{starkit} \citep{wolfgang_kerzendorf_2015_28016}}}

\newglossaryentry{pyne}{name=PYNE, text=\textsc{pyne}, description= PYNE code, first = {\textsc{pyne} \citep{Scopatz2012a}}}
\newglossaryentry{multinest}{name=MULTINEST, text=\textsc{MultiNest}, description=MultiNest, first={\textsc{MultiNest} \citep{2009MNRAS.398.1601F}}}
\newglossaryentry{wsynphot}{name=WSYNPHOT, text=\textsc{wsynphot}, description=Wsynphot, first={\textsc{wsynphot}\protect\footnote{\protect\url{https://github.com/wkerzendorf/wsynphot}}}}
\newglossaryentry{specutils}{name=SPECUTILS, text=\textsc{specutils}, description=specutils, first={\textsc{specutils} \protect\footnote{\protect\url{https://github.com/astropy/specutils}}}}
\newglossaryentry{ads}{name=ADS ,description=ADS, first={NASA Astrophysics Data System (ADS) \citep{2000A&AS..143...41K}}}

\newglossaryentry{2mass}{name=2MASS,description={Two Micron All Sky Survey \citep{2006AJ....131.1163S}}, first={Two Micron All Sky Survey \citep{2006AJ....131.1163S}}}
\newglossaryentry{wiserep}{name=\textsc{WISeREP}, description={Weizmann Interactive Supernova data REPository \citep{2006AJ....131.1163S}}, first={\textsc{WISeREP} \citep{2012PASP..124..668Y}}}
\newglossaryentry{nomad}{name=NOMAD,first={Naval Observatory Merged Astrometric Dataset \citep[NOMAD; ][]{2005yCat.1297....0Z}}, description={Naval Observatory Merged Astrometric Dataset}}
\newglossaryentry{wifes}{name=WIFES, text=\textsc{WiFeS}, first={\textsc{WiFeS} \citep{2007Ap&SS.310..255D}},  description={Wide Field Spectrograph - \gls{ifu} mounted on the 2.3\,m telescope at Siding Spring Observatory}}
\newglossaryentry{scp}{name=SCP,description={Supernova Cosmology Project, led by Saul Perlmutter}, first={Supernova Cosmology Project (SCP)}}
\newglossaryentry{hzsns}{name=HZSNS,description={High Z Supernova Search, led by Brian Schmidt}, first={High Z Supernova Search (HZSNS)}}
\newglossaryentry{vlt}{name=VLT,description={Very Large Telescope located on Cerro Paranal (Chile)}, first={Very Large Telescope (VLT)}}
\newglossaryentry{flames}{name=FLAMES,description={Multi-object, intermediate and high resolution spectrograph mounted on the  \gls{vlt}}}
\newglossaryentry{hires}{name=HIRES, description={High Resolution Echelle Spectrometer mounted on the Keck Telescope}, first={High Resolution Echelle Spectrometer \citep[HIRES;][]{1994SPIE.2198..362V}}}
\newglossaryentry{lris}{name=LRIS,description={Low Resolution Imaging Spectrometer mounted on the Keck Telescope}, first={Low-Resolution Imaging Spectrometer \citep[LRIS;][]{Oke95}}}

\newglossaryentry{decam}{name=DECam, description={DECam is a high-performance, wide-field CCD imager mounted at the prime focus of the Blanco 4-m telescope at \gls{ctio}.}, first={Dark Energy Camera \citep[DECam; ][]{2012PhPro..37.1332D,2015AJ....150..150F}}}

\newglossaryentry{essence}{name=ESSENCE,description={The `Equation of State: SupErNovae trace Cosmic Expansion' project \citep[ESSENCE;][]{2002AAS...201.7809G}}, first={`The Equation of State: SupErNovae trace Cosmic Expansion' \citep[ESSENCE;][]{2002AAS...201.7809G}}}
\newglossaryentry{ifu}{name=IFU,description={Optical instrument combining spectrographic and imaging capabilities, used to obtain spatially resolved spectra}, first={Integral Field Unit (IFU)}, firstplural={Integral Field Units (IFUs)}}

\newglossaryentry{besancon}{name=Besan\c{c}on Model, description={Model of stellar population synthesis of the Galaxy, including kinematics.}}

\newglossaryentry{int}{name=INT,description={Isaac Newton 2.5\,m Telescope}, first={Isaac Newton 2.5\,m Telescope (INT)}}
\newglossaryentry{iau}{name=IAU,description={International Astronomical Union}, first={IAU}}
\newglossaryentry{chandra}{name=Chandra,description={Chandra \xray\ Observatory (space-based)}}
\newglossaryentry{hst}{name=HST,description={Hubble Space Telescope}}
\newglossaryentry{hst.wfpc2}{name=WFPC2,description={Wide-Field Planetary Camera 2 mounted on the \gls{hst}}, first={Wide-Field Planetary Camera 2 (WFPC2)}}
\newglossaryentry{hst.acs}{name=ACS,description={Advanced Camera for Surveys mounted on the \gls{hst}}, first={Advanced Camera for Surveys (ACS)}}
\newglossaryentry{hst.wfc3}{name=WFC3,description={Wide-Field Camera 3 mounted on the \gls{hst}}, first={Wide-Field Camera 3 (WFC3)}}
\newglossaryentry{hst.cte}{name=CTE, description={charge transfer efficiency (CTE)}, first={charge transfer efficiency \citep[CTE; see ][for a description]{2009acs..rept....1C}}}

\newglossaryentry{snls}{name=SNLS,description={Supernova Legacy Survey \citep{2003AAS...203.8209P}}, first={Supernova Legacy Survey \citep[SNLS;][]{2003AAS...203.8209P}}}
\newglossaryentry{dass}{name=DASS, description={Digitized Astronomy Supernova Survey \citep{1975PASP...87..565C}}, first={Digitized Astronomy Supernova Survey \citep[DASS;][]{1975PASP...87..565C}}}
\newglossaryentry{bait}{name=BAIT, description={Berkley Automatic Imaging Telescope \citep{1993PASP..105.1164R}}, first={Berkley Automatic Imaging Telescope \citep[BAIT;][]{1993PASP..105.1164R}}}
\newglossaryentry{kait}{name=KAIT, description={Katzman Automatic Imaging Telescope \citep{2001ASPC..246..121F}}, first={Katzman Automatic Imaging Telescope \citep[KAIT;][]{2001ASPC..246..121F}}}
\newglossaryentry{loss}{name=LOSS, description={Lick Observatory Supernova Search  \citep{2000AIPC..522..103L}}, first={Lick Observatory Supernova Search \citep[LOSS;][]{2000AIPC..522..103L}}}
\newglossaryentry{ctss}{name=CTSS,description={Cal\'{a}n/Tololo Supernova Survey \citep{1993AJ....106.2392H}}, first={Cal\'{a}n/Tololo supernova survey \citep[CTSS;][]{1993AJ....106.2392H}}}
\newglossaryentry{ctio}{name= CTIO, description={Cerro Tololo Inter-American Observatory}, first={Cerro Tololo Inter-American Observatory (CTIO)}}
\newglossaryentry{ptf}{name=PTF, description={Palomar Transient Factory \citep{2009PASP..121.1334R}}, first={Palomar Transient Factory \citep[PTF;][]{2009PASP..121.1334R}}}
\newglossaryentry{batse}{name=BATSE, description={Burst and Transient Source Experiment mounted on the Compton Gamma Ray Observatory}, first={Burst and Transient Source Experiment (BATSE)}}
\newglossaryentry{bepposax}{name=BeppoSAX, description={\xray\ satellite named in honor of Giuseppe "Beppo" Occhialini}}
\newglossaryentry{rosat}{name=ROSAT, description={short for R\"{o}ntgensatellit}, first={ROSAT}}
\newglossaryentry{hete2}{name=HETE2, description={High Energy Transient Explorer}, first={High Energy Transient Explorer (HETE)}}
\newglossaryentry{ska}{name=SKA, description={Square Kilometre Array}, first={Square Kilometre Array (SKA)}}

\newglossaryentry{gnirs}{name=GNIRS, description={Gemini Near InfraRed Spectrograph mounted on the Gemini North Telescope}}
\newglossaryentry{gmosn}{name=GMOS, description={Gemini Multi Object Spectrograph mounted on the
 Gemini North Telescope}, first={GMOS \citep[Gemini Multi Object Spectrograph;][]{2004PASP..116..425H}}}
\newglossaryentry{swift}{name=Swift, description={Swift Gamma-Ray Burst Mission}}
\newglossaryentry{vla}{name=VLA, description={Very Large Array radio telescope located in North America}, first={Very Large Array (VLA)}}
\newglossaryentry{evla}{name=EVLA, description={Extended Very Large Array radio telescope located in North America}, first={Extended Very Large Array (EVLA)}}
\newglossaryentry{sdss}{name=SDSS, description={Sloan Digital Sky Survey}}
\newglossaryentry{dss}{name=DSS, description={Digitized Sky Survey}}
\newglossaryentry{skymapper}{name=SkyMapper, description={SkyMapper telescope \citep{2007PASA...24....1K}}, first={SkyMapper \citep{2007PASA...24....1K}}}
\newglossaryentry{panstarrs}{name=PanSTARRS, description={Panoramic Survey Telescope \& Rapid Response System \citep{2004SPIE.5489...11K}}, first={Panoramic Survey Telescope \& Rapid Response System \citep[PanSTARRS;][]{2004SPIE.5489...11K}}}
\newglossaryentry{ps1dr1}{name=PS1~DR1, description={Panoramic Survey Telescope \& Rapid Response System \citep{2004SPIE.5489...11K} }, first={Panoramic Survey Telescope \& Rapid Response System \citep[PanSTARRS;][]{2004SPIE.5489...11K} DR1}}

\newglossaryentry{lsst}{name=LSST, description={Large Synoptic Survey Telescope}, first={Large Synoptic Survey Telescope \citep[LSST;][]{2006AAS...209.8604P}}}
\newglossaryentry{ppmxl}{name=PPMXL, description={PPMXL Catalog of Positions and Proper Motions on the ICRS \citep{2010AJ....139.2440R}}}
\newglossaryentry{gaia}{name=GAIA, description={Global Astrometric Interferometer for Astrophysics \citep{2001A&A...369..339P}}, first={Global Astrometric Interferometer for Astrophysics \citep[GAIA;][]{2001A&A...369..339P}}}
\newglossaryentry{ligo}{name=LIGO, description={Laser Interferometer Gravitational Wave Observatory}, first={Laser Interferometer Gravitational Wave Observatory \citep[LIGO;][]{1992Sci...256..325A}}}
\newglossaryentry{aligo}{name=Advanced LIGO, description={Advanced LIGO}, sort=ligo2}
\newglossaryentry{lisa}{name=LISA, description={Laser Interferometer Space Antenna \citep{1994ESAJ...18..219J}}, first={Laser Interferometer Space Antenna \citep[LISA;][]{1994ESAJ...18..219J}}}
\newglossaryentry{ccd}{name=CCD,description={Charged Coupled Device}, first={charged coupled device (CCD)}, firstplural={charged coupled devices (CCDs)}}

\newcommand{\sn}[2]{SN~#1#2\xspace}

\newglossaryentry{irc}{name=IRC, text={IRC}, description={infrared catastrophe}, first={infrared catastrophe \citep[IRC;][]{1980PhDT.........1A}}}

\newglossaryentry{sn}{name=Supernova, text={SN}, plural={SNe}, description={exploding star}, nonumberlist=true, first={supernova (SN)}, firstplural={supernovae (SNe)}}
\newglossaryentry{snia}{name=Type~Ia (SN~Ia), text={SN~Ia}, description={Thermonuclear explosion of a white dwarf - spectra show no hydrogen but a strong silicon line},first={Type~Ia supernova (SN~Ia)}, firstplural={Type Ia supernovae (SNe~Ia)}, plural={SNe~Ia}, parent=sn, nonumberlist=true}
\newcommand{\sneia}{\glspl*{snia}\xspace}

\newglossaryentry{branchnormal}{name={branch-normal}, text=\textit{Branch-normal}, description={Large homogeneous class of Type Ia Supernovae, defined in \citet{1993AJ....106.2383B}}, first={\textit{Branch-normal} SNe Ia \citep{1993AJ....106.2383B}}, parent=snia}
\newglossaryentry{91t}{name={91T-like}, description={Luminous class of Type Ia supernovae similar to \sn{1991}{T} \citep{1992AJ....103.1632P}} , first={91T-like}, parent=snia}
\newglossaryentry{91bg}{name={91bg-like}, description={Faint class of Type Ia supernovae similar to \sn{1991}{bg} \citep{1992AJ....104.1543F}}, first={91bg-like}, parent=snia}
\newglossaryentry{02cx}{name={02cx-like}, description={Peculiar class of Type Ia supernovae similar to \sn{2002}{cx} \citep{2003PASP..115..453L}}, first={02cx-like \sneia\ \citep{2003PASP..115..453L}}, parent=snia}

\newglossaryentry{snibc}{name=Type~Ib/c, text={SN~Ib/c}, description={Collapse of the core of a massive star -  spectrum shows no hydrogen and no silicon line},first={Type~Ib/c supernova (SN~Ib/c)}, firstplural={Type~Ib/c supernovae (SNe~Ib/c)}, plural={SNe~Ib/c}, parent=sn}

\newglossaryentry{snib}{name=Type~Ib, text={SN~Ib}, description={Spectrum shows no hydrogen and no silicon, but helium line},first={Type Ib supernova (SN~Ib)}, firstplural={Type~Ib supernovae (SNe~Ib)}, plural={SNe~Ib}, parent=snibc}

\newglossaryentry{snic}{name=Type~Ic, text={SN~Ic}, description={Spectrum shows no hydrogen, no silicon and no helium line},first={Type~Ic supernova (SN~Ic)}, firstplural={Type~Ic supernovae (SNe~Ic)}, plural={SNe~Ic}, parent=snibc}

\newglossaryentry{snii}{name=Type~II, text={SN~II}, description={Collapse of the core of a massive star - spectrum shows strong hydrogen line},first={Type~II supernova (SN~II)}, firstplural={Type~II supernovae (SNe~II)}, plural={SNe~II}, parent=sn}

\newglossaryentry{sniib}{name=Type~IIb, text={SN~IIb}, description={Spectrum shows hydrogen and helium lines},first={Type~IIb supernova (SN~IIb)}, firstplural={Type~IIb supernovae (SNe~IIb)}, plural={SNe~IIb}, parent=snii}

\newglossaryentry{sniip}{name=Type~II~Plateau (Type IIP), text={SN~IIP}, description={Lightcurve shows plateau},first={Type~IIP supernova (SN~IIP)}, firstplural={Type~II Plateau supernovae \citep[SNe~IIP;][]{1979A&A....72..287B}}, plural={SNe~IIP}, parent=snii}

\newglossaryentry{sniil}{name=SN~II~Linear, text={SN~IIL}, description={Lightcurve shows no plateau, but linear decline},first={Type~IIL supernova (SN~IIL)}, firstplural={Type~II~Linear supernovae \citep[SNe~IIL;][]{1990MNRAS.244..269S}}, plural={SNe~IIL}, parent=snii}

\newglossaryentry{sniin}{name=Type II narrow-lined (Type IIn), description={Spectrum shows narrow lines},first={Type~II~narrow-lined supernova (SN IIn)}, firstplural={Type~IIn supernovae (SNe~IIn)}, plural={SNe~IIn}, parent=snii}

\newglossaryentry{snr}{name=Remnant (SNR), text=SNR, description={Remnant left visible post-explosion}, first={supernova remnant (SNR)}, firstplural={supernova remnants (SNRs)}, parent=sn}

\newglossaryentry{dtd}{name=DTD,description={delay time distribution - expected supernova rate over time after a brief outburst of starformation},first={delay time distribution (DTD)}, firstplural={delay time distributions (DTDs)}, plural=DTDs}

\newglossaryentry{hvg}{name=HVG,description={high velocity gradient - Type Ia supernovae with a fast evolution of photospheric velocity},first={high velocity group (HVG)}, firstplural={high velocity groups (HVGs)}, plural=HVGs, parent=snia}

\newglossaryentry{lvg}{name=LVG,description={low velocity gradient - Type Ia supernovae with a slow evolution of photospheric velocity},first={low velocity group (LVG)}, firstplural={low velocity groups (LVGs)}, plural=LVGs, parent=snia}

\newglossaryentry{wd}{name=white dwarf (WD), text=WD, description={White Dwarf - extremely dense stellar remnant}, first={white dwarf (WD)}}
\newglossaryentry{onemgwd}{name= Oxygen/Neon (ONe), text={ONe-WD},description={Oxygen/Neon White Dwarf}, first={oxygen/neon White Dwarf (ONe-WD)}, parent=wd}
\newglossaryentry{cowd}{name=carbon/oxygen (CO), text={CO-WD}, description={carbon/oxygen white dwarf}, first={carbon/oxygen white dwarf (CO-WD)}, firstplural = {carbon/oxygen white dwarfs (CO-WDs)}, parent=wd}

\newglossaryentry{sds}{name=SD-Scenario,description={single-degenerate scenario (single white dwarf accreting from non-degenerate companion)}, first={single-degenerate scenario (SD-scenario)}}

\newglossaryentry{dds}{name=DD-Scenario, description={double degenerate scenario (merging of two white dwarfs)}, first={double-degenerate scenario (DD-scenario)}}

\newglossaryentry{sss}{name=SSS, text={supersoft \xray\ source}, description={supersoft \xray\ source - believed to be emitted by nuclear fusion on a white dwarf's surface}}

\newglossaryentry{amcvn}{name=AM CVn, description={AM Canum Venaticorum star \citep[white dwarf accreting hydrogen poor matter from a companion star; see ][]{2005ASPC..330...27N}}}

\newglossaryentry{rlof}{name=RLOF, description={Roche Lobe Overflow (see \citet{1971ARA&A...9..183P} for a more detailed description)}, first={Roche-lobe overflow (RLOF)}}

\newglossaryentry{mchan}{name={Chandrasekhar mass~}, text={Chandrasekhar~mass}, symbol={\ensuremath{M_\textrm{Chan}}}, plural={Chandrasekhar~masses}, description={Mass when the core of a star collapses due to insufficient degeneracy pressure - for a white dwarf $\approx1.38\,M_\odot$ see \citet{1931ApJ....74...81C}}, first={Chandrasekhar~mass \citep[$M_\textrm{Chan}=1.38\,M_\odot$;][]{1931ApJ....74...81C}}, sort=mchan}

\newglossaryentry{w7}{name={W7 model},description={W7 model \citep{1984ApJ...286..644N}},first = {W7 model \citep{1984ApJ...286..644N}}}

\newglossaryentry{ew}{name=Equivalent Width, text={EW}, description={width of a rectangle that has the same area as a spectral line when taken to zero flux}, first={equivalent width (EW)}, firstplural={equivalent widths (EWs)}}
\newglossaryentry{agb}{name=AGB,description={Asymptotic Giant Branch}, first={Asymptotic Giant Branch (AGB)}}
\newglossaryentry{cmb}{name=CMB,description={Cosmic Microwave Background}}
\newglossaryentry{csm}{name=CSM,description={Circumstellar Medium}, first={circumstellar medium (CSM)}}
\newglossaryentry{csi}{name=CSI,description={Circumstellar Interaction}, first={circumstellar interaction (CSI)}}
\newglossaryentry{ism}{name=ISM,description={Interstellar Medium}, first={interstellar medium (ISM)}}
\newglossaryentry{ige}{name=IGE,description={Iron Group Element}, first={iron group element (IGE)}, firstplural={iron group elements (IGEs)}}
\newglossaryentry{epm}{name=EPM,description={Expanding Photosphere Method \citep{1974ApJ...193...27K}}, first={Expanding Photosphere Method (EPM)}}
\newglossaryentry{aic}{name=AIC,description={Accretion Induced Collapse}, first={accretion induced collapse (AIC)}}
\newglossaryentry{ime}{name=IME,description={Intermediate Mass Element}, first={intermediate mass element (IME)}, firstplural={intermediate mass elements (IMEs)}}
\newglossaryentry{h0}{name=\ensuremath{H_0},description={Hubbles constant}}
\newglossaryentry{nse}{name=NSE,description={Nuclear Statistical Equilibrium}, first={nuclear statistical equilibrium (NSE)}}
\newglossaryentry{cdm}{name=CDM,description={Cold Dark Matter}}
\newglossaryentry{grb}{name=GRB,description={Gamma Ray Burst}, first={Gamma Ray Burst (GRB)}, firstplural={Gamma Ray Bursts (GRBs)}}
\newglossaryentry{xps}{name=XPS, description={X-ray point source}, first={X-ray point source (XPS)}, firstplural={X-ray point sources (XPS)}}
\newglossaryentry{donor}{name=donor,description={non-degenerate companion in the \gls{sds}}}
\newglossaryentry{mainsequence}{name=main sequence,description={main sequence star}}
\newglossaryentry{redgiant}{name=red giant,description={red giant star}}
\newglossaryentry{mlcs}{name=MLCS,description={Multicolor Light Curve Shape method \citep[MLCS;][]{1996ApJ...473...88R}}, first={Multicolor Light-Curve Shape method \citep[MLCS;][]{1996ApJ...473...88R}}}
\newglossaryentry{rsoph}{name=RS~Ophiuci ,description={white dwarf accreting from a red giant - assumed progenitor of the \gls{sds}}, sort=rsoph}
\newglossaryentry{usco}{name=U~Scorpii,description={white dwarf accreting from a main sequence star - assumed progenitor of the \gls{sds}}, sort=usco}
\newglossaryentry{rcw86}{name=RCW~86,description={supernova remnant sometimes associated with \sn{185}{}}, sort=rcw86}
\newglossaryentry{casa}{name=Cas~A,description={Cassiopeia A supernova remnant - probably a \gls{snib} event}}
\newglossaryentry{cepheid}{name=Cepheid,description={very luminous variable star with a strong luminosity period relationship}}
\newglossaryentry{urca}{name=Urca, text=\textit{Urca}, description={process predominatly contributing to cooling in stars. The \textit{Urca} process consists of alternating electron-capture and $\beta^{-}$ decay of two nuclei pairs.},sort=urca}
\newglossaryentry{alphacen}{name=Alpha Centauri,description={one of the brightest stars in the night sky and a close binary}}
\newglossaryentry{pcygni}{name={P Cygni}, text={P Cygni},description={a hypergiant luminous blue variable with strong winds. Often referred to as a description for their line profiles showing a emission peak at the rest wavelength of the line and a blue-shifted absorption trough.}}

\newglossaryentry{teff}{name={effective temperature~}, text={effective temperature}, symbol={\ensuremath{T_\textrm{eff}}}, description={Temperature of a blackbody emitting the same total energy}, sort=teff}

\newglossaryentry{logg}{name={surface gravity~}, text={surface gravity}, symbol={\ensuremath{\textrm{log}\,g}}, description={gravity at the surface of a star}, sort=logg}
\newglossaryentry{feh}{name={metallicity~}, text={metallicity}, symbol=\textrm{[Fe/H]},description={iron abundance relative to the sun}, sort=feh}

\newglossaryentry{texp}{name={time since explosion~}, text={time since explosion}, text={time since explosion}, symbol={\ensuremath{t_{\rm exp}}},description={time since explosion (measured in days)}, sort=texp, first={time since explosion (\ensuremath{t_{\rm exp}})}}

\newglossaryentry{lmc}{name=LMC,description={Large Magellanic Cloud}, first={Large Magellanic Cloud (LMC)}, sort=lmc}
\newglossaryentry{smc}{name=SMC,description={Small Magellanic Cloud}, sort=smc}
\newglossaryentry{z}{name=\ensuremath{z},description={redshift}, sort=z}

\newglossaryentry{stats.pdf}{name=PDF, description={Probability Density Function}, first={Probability Density Function}}

\makeglossaries

\graphicspath{{./}{figures/}}

\submitjournal{ApJ}

\shorttitle{TARDIS SN1994I}
\shortauthors{Williamson et al.}

\begin{document}

\title{Modelling Type Ic Supernovae with \textsc{tardis}: Hidden Helium in SN1994I?}

\author{Marc Williamson}
\affil{New York University}

\author{Wolfgang Kerzendorf}
\affiliation{Michigan State University}

\author{Maryam Modjaz}
\affiliation{New York University}

\begin{abstract}
Supernovae (SNe) with photospheric spectra devoid of Hydrogen and Helium features are generally classified as Type Ic SNe (SNe Ic). However, there is ongoing debate as to whether Helium can be hidden in the ejecta of SNe Ic (that is, Helium is present in the ejecta, but produces no obvious features in the spectra). We present the first application of the fast, 1-D radiative transfer code \textsc{tardis} to a SN Ic, and we investigate the question of how much Helium can be hidden in the outer layers of the SN Ic ejecta. We generate \textsc{tardis} models for the nearby, well-observed, and extensively modeled SN Ic 1994I, and we perform a code comparison to a different, well-established Monte Carlo based radiation transfer code. The code comparison shows that \textsc{tardis} produces consistent synthetic spectra for identical ejecta models of SN1994I. In addition, we perform a systematic experiment of adding outer He shells of varying masses to our SN1994I models. We find that an outer He shell of only $0.05M_{\odot}$ produces strong optical and NIR He spectral features for SN1994I which are not present in observations, thus indicating that the SN1994I ejecta is almost fully He deficient compared to the He masses of typical He-rich SN progenitors. Finally we show that the He I $\lambda$20851 line pseudo equivalent width of our modeled spectra for SN1994I could be used to infer the outer He shell mass which suggests that NIR spectral follow-up of SNe Ic will be critical for addressing the hidden helium question for a statistical sample of SNe Ic.
\end{abstract}


\section{Introduction} \label{sec:intro}
Stripped-Envelope (SE) supernovae (SNe) are the core collapse explosions of massive $(>8M_{\odot})$ stars that have lost part or all of their outermost Hydrogen and Helium layers \citep{clocchiatti1997sn}. In particular, SNe with photospheric spectra lacking Hydrogen and Helium features are classified as type Ic \citep[SNe~Ic,][]{filippenko1993type,modjaz19_review}. There are many potential underlying causes for the lack of both Hydrogen and Helium features in SNe Ic spectra. Stripping due to binary interaction \citep{podsiadlowski92,yoon_ibc_prog_rev}, line-driven winds for massive single progenitors \citep{crowther07,smartt09_rev}, and homogeneous chemical evolution due to rotation \citep[only for H,][]{maeder87,langer2012presupernova} are all possible mechanisms for removing the outer layers of SN Ic progenitors \citep[for more review of SESNe progenitor scenarios see][]{smartt09_rev,yoon_ibc_prog_rev}.

In the last decade, the question of hidden He has arisen (i.e. He present in the SN ejecta without producing observable spectral absorption features). Complex radiative transfer simulations like \gls{cmfgen} show that it is possible to hide up to the entire outer He layer \citep{dessart12} if there is low mixing of $^{56}$Ni. In this case, optical He I lines are not present because they require non-thermal excitation by fast electrons which are produced through interactions with $\gamma$-rays from $^{56}$Ni and $^{56}$Co decay \citep{lucy91}. However, there is also evidence for particular SNe Ic like SN1994I where radiative transfer simulations show that optical He I lines do appear when an outer He layer of mass as low as $\approx 0.10M_{\odot}$ is present \gls{hach12}. Understanding how much He can be hidden in SNe Ic ejecta is crucial for understanding the SN Ic progenitor scenarios. Specifically, binary system progenitor models like those presented in \cite{dessart2020supernovae} and \cite{woosley2020ibc} can be tuned to produce an appropriate amount of stripping, and single Wolf-Rayet star progenitor models like those presented by \cite{yoon2017towards} can be tuned to have appropriate mass loss rates. Multiple theoretical studies have been published that discuss hidden helium focusing on different progenitor and explosion parameters like binary progenitor system parameters \citep{dessart15} and the degree of $^{56}$Ni mixing \citep{dessart12}. However there are very few modeling efforts in the literature dedicated to observed SNe Ic \citep{swartz87M,iwamoto87ef,mazzali87ef,iwamoto94, hachinger94Imodel} and none perform a systematic investigation to constrain the possible amount of hidden helium. There is a clear need for such hidden helium focused modeling efforts for observed SNe Ic in order to understand the diversity of the SN Ic class.

SN~1994I is one of the best observed SNe Ic \citep{filippenko1995type, clocchiatti96} since it occurred in the nearby galaxy M51. Recent analysis of the light curve \citep{drout2011first,bianco_vband_conv} and spectra \citep{modjaz_icbl,Williamson19} has shown that SN1994I may actually be relatively atypical compared to other SNe Ic. However, SN1994I is an excellent choice for modelling due to the large number of modelling attempts in the literature \citep{Baron94Imodel, sauer94Imodel, hachinger94Imodel, parrent94Imodel} and high quality observational data \citep{filippenko1995type, clocchiatti96}. Despite the wealth of papers in the literature modelling SN~1994I, there has not been a systematic investigation to constrain the presence or amount of hidden He.  \cite{Baron94Imodel} model SN1994I using \gls{phoenix}, but their models are relatively insensitive to He abundance, so the He mass cannot be inferred. \cite{sauer94Imodel} use the Monte Carlo (MC) radiative transfer approach \citep{1985apj...288..679a, lucyabbott1993, 1993a&a...279..447m, lucy99, 2002a&a...384..725l, 2003a&a...403..261l,mazzali00}, but their simulations do not include the treatment of the important  non-thermal fast electrons, which are known to affect the excitation state of He \citep{lucy91}. More recently, \gls{hach12} computed models of SN~1994I which included a treatment of the important non-thermal processes. They compute a model sequence between SN~1994I and the SN IIb 2008ax \citep{pastorello2008type}, but this work lacks a systematic exploration of how much He can be contained in the SN~1994I ejecta. There is a clear need for such a systematic investigation into the possible presence and amount of He in the SN~1994I ejecta using a fast radiation transport code capable of treating the non-thermal processes.

In this paper, we perform a systematic set of spectral synthesis simulations using the open-source radiation transport code \gls{tardis}, which includes a treatment for non-thermal processes affecting He \citep{boyle_tardis}, in order to place a strong upper limit on the amount of hidden He that can exist in the ejecta for SN~1994I.  SN~1994I is a good candidate for understanding the hidden He question because there are relatively strong constraints on its progenitor scenario due to hydrodynamical modeling that simultaneously fit the monochromatic light curves \citep{iwamoto94}. \cite{iwamoto94} rule out the single Wolf-Rayet star progenitor scenario due to the fast decline of the SN~1994I light curve relative to other SNe Ic \citep{bianco_vband_conv}. Further modeling by \cite{van2016constraints} narrows down the three binary system progenitor scenarios presented by \cite{iwamoto94} to SN~1994I resulting from the explosion of a low mass C+O star with a low or intermediate mass main sequence binary companion. Both low \citep{iwamoto94} and high \citep{sauer94Imodel} degrees of $^{56}$Ni mixing have been proposed for the SN~1994I ejecta, but for the purposes of establishing an upper limit to the amount of hidden He, we adopt the low mixing CO21 model presented by \cite{iwamoto94} since increased mixing will produce stronger lines in the spectra, for the same He mass. Therefore, a low $^{56}$Ni mixing model is the most conservative choice for testing the maximum amount He that can be hidden in the SN~1994I ejecta.

As we will show with our \gls{tardis} models of SN~1994I, distinct He I absorption features appear in the optical and near infrared (NIR) spectra with an outer He shell of mass $0.05M_{\odot}$. In addition, we will show that not only is the He I 20851 line sensitive to the existence of trace amounts of He, but how its pseudo-equivalent width (pEW) increases with increasing He outer shell mass. 

In Section \ref{sec:methods} we discuss the fast, 1D, time independent, open-source\footnote{\url{https://github.com/tardis-sn/tardis}} radiation transfer code \gls{tardis} used for our simulations. We present the methodology of our code comparison to a similar Monte Carlo based radiation transport code (\gls{hach12}) and the methodology for our detailed hidden He investigation. The results of our code comparison and hidden He investigation are presented in Section \ref{sec:results}, and we conclude in Section \ref{sec:summary}.

\section{Methods} \label{sec:methods}
In Section \ref{sec:methods_tardis} we present details of the code \gls{tardis}, the radiative transfer code used for the simulations in this work. We discuss specific parameter choices for our code comparison with \gls{hach12} in Section \ref{sec:methods_cc}, and we present our method for investigating the presence of hidden helium in SN1994I models in Section \ref{sec:methods_hh}.

\subsection{\textsc{TARDIS}} \label{sec:methods_tardis}
\gls{tardis} is an MC radiative transfer code based on the approach presented in \cite{1985apj...288..679a, lucyabbott1993, 1993a&a...279..447m, lucy99, 2002a&a...384..725l, 2003a&a...403..261l}. For a given SN ejecta model, the user must provide the abundance and density structure as input, and \gls{tardis} self-consistently solves for the ionization and excitation state of the plasma by propagating photon packets through the ejecta. \gls{tardis} assumes homologous expansion, which is a reasonable assumption for the spectra we are modelling, since the ejecta stratification structure is mostly fixed by 5 days post explosion \citep{tsang_homology}, and in this work we only model spectra between 16 and 40 days post explosion. In this early photospheric phase of evolution, \gls{tardis} approximates the supernova as an optically thick core emitting a blackbody continuum, surrounded by more transparent outer shells which are responsible for absorption features. MC photon packets are assumed to originate from a blackbody profile at a constant velocity in the ejecta model. Photon packets accrue optical depth as they propagate through the ejecta and probabilistically experience either electron scattering or atomic line  transitions. For the simulations presented in this paper, we use the Kurucz atomic dataset (i.e. line list) for calculating the bound-bound transitions \citep{kuruczbell1995} with the H and He lines taken from the \gls{chianti} database. \gls{tardis} supports a full macroatom implementation of atomic transitions \citep{lucy_macroatom} as well as a modified version of macroatom called downbranch, in addition to pure scattering. While above we describe the general features of \gls{tardis}, we proceed below to give more detailed information on the two settings in which we run \gls{tardis} for SN1994I.

\subsection{Code Comparison} \label{sec:methods_cc}
This paper is the first application of \gls{tardis} to SNe Ic, so we conduct a code comparison to the similarly-designed MC radiative transfer code used by \gls{hach12} to model SN~1994I. Other radiative transfer codes have been used to model SN~1994I, but algorithmic differences between these codes and \gls{tardis} would greatly obfuscate the code comparison. In particular, the \gls{phoenix} code used by \cite{Baron94Imodel} is time dependent (unlike \gls{tardis}) and the MC code used by \cite{sauer94Imodel} does not include a treatment for the non-thermal excitation of Helium (unlike \gls{tardis}). The code comparison to \gls{hach12} allows us to test the analytic approximation of \cite{boyle_tardis} that \gls{tardis} uses to incorporate the non-thermal effects on He from fast electrons against the full non-local thermodynamic equilibrium (NLTE) treatment used in \gls{hach12}. We have obtained the low He mass ($\sim 0.01M_{\odot}$) models of SN~1994I used in \gls{hach12} (see \cite{hachinger94Imodel} Fig. 5) for the SN~1994I spectra at phases 16, 22, 30, and 40 days post explosion (priv. comm. Hachinger). For the code comparison, we use the \gls{hach12} input models to generate \gls{tardis} synthetic spectra and compare the \gls{tardis} results to their model spectra presented in \gls{hach12}. In this section, we describe the \gls{tardis} settings chosen to facilitate an equal comparison to \gls{hach12}, and the most important parameter choices are recorded in Table \ref{tab:hach_direct_table}. The \gls{tardis} configuration and model files are publicly available.

\begin{deluxetable}{CC}[ht!]
    \tablecaption{\gls{tardis} Parameter Settings for Code Comparison\label{tab:hach_direct_table}}
    \tablecolumns{2}
    \tablenum{1}
    \tablewidth{0pt}
    \tablehead{\colhead{\gls{tardis} Parameter} &\colhead{Setting}}
    \startdata
        \text{Ionization} & \texttt{nebular} \\
        \text{Excitation} & \texttt{dilute-lte} \\
        \text{Radiative Rate} & \texttt{dilute-blackbody} \\
        \text{Line Interaction} & \texttt{downbranch} \\
        \text{Number of Iterations} & 50 \\
        \text{Number of Packets} & $1.0\times10^{5}$ \\
        \text{Helium Treatment} & \texttt{recomb-nlte}
    \enddata
\end{deluxetable}
We run \gls{tardis} using the \texttt{nebular} Ionization mode and the \texttt{dilute-lte} excitation mode. This is a first order departure from a standard Saha-Boltzmann LTE treatment that accounts for dilution of the radiation field. Although \gls{tardis} supports a full macroatom implementation of atomic transitions during bound-bound processes, we use \texttt{downbranch} mode for consistency with \gls{hach12}. The \texttt{dilute-blackbody} mode for radiative rate enforces that \gls{tardis} calculates radiative rates in the same way as \gls{hach12}. One minor difference between \gls{tardis} and the code used in \gls{hach12} involves the treatment of energy deposition due to gamma rays. Fast Compton electrons are produced due to gamma rays from the $^{56}$Ni decay, and it has been shown that these non-thermal electrons have an effect on the ionization and excitation state of He \citep{lucy91}. \gls{hach12} use a light curve modeling code to simulate the creation and transport of gamma rays and incorporate this information into their NLTE treatment. While \gls{tardis} does not explicitly simulate gamma rays, it supports the \texttt{recomb-nlte} He treatment \citep{boyle_tardis}, which is an approximation for Helium based on the simulations in  \gls{hach12} that show the He I ground state population is negligible due to non-thermal effects. The \texttt{recomb-nlte} mode treats the He I excited states as being in dilute LTE with the He II ground state, and calculates the He II excited states and He III ground state relative to the He II ground state. 

We use the same density, abundance, and radiative temperature structure as \gls{hach12}. By forcing \gls{tardis} to use the $T_{rad}$ profile from \gls{hach12}, we are using \gls{tardis} as an opacity calculator. The density profile is the result of the CO21 hydrodynamical simulation of a low mass C+O stellar core explosion \citep{iwamoto94}. We adopt the CO21 model in order to facilitate an equal comparison with \gls{hach12}, which used the same model. However we note that there are many more recent stellar evolution models of stripped stars that may give rise to SESNe \citep{yoon2017towards,laplace2020expansion,dessart2020supernovae,woosley2020ibc}.

\subsection{Hidden Helium Investigation} \label{sec:methods_hh}
Despite the large number of high quality observed spectra \citep{filippenko1995type,clocchiatti96} and multiple synthetic models for SN~1994I \citep{Baron94Imodel,sauer94Imodel,hachinger94Imodel} in the literature, there has not been a systematic modelling investigation to determine the amount of Helium that could be hidden in the SN~1994I ejecta. Constraining the Helium content of SN~1994I and other SNe Ic will inform stellar evolution simulations of the progenitors of these explosions. Most massive stars are formed in binaries \citep{sana_massive_binarity}, and binary systems in particular are most likely responsible for the stripping necessary to produce SNe Ic \citep{dessart11,smith11_snfrac,langer12,dessart2020supernovae,woosley2020ibc,schneider2020}. Hidden He can affect parameters of the binary system like companion mass and type of mass transfer required to strip the right amount of He from the host \citep{dewi_he_star}. In addition, given recent advances in the understanding of Wolf-Rayet star mass-loss rates, it is possible that low mass single He stars could be the progenitors for a fraction of SNe Ic \citep{yoon2017towards}. In this case, hidden He would affect our understanding of the degree of mass loss required to produce an explosion like SN~1994I.

In this section, we describe the systematic process used in this work to determine an upper limit on the mass of the outer He shell for SN1994I. We use the same \gls{tardis} settings as described in Section \ref{sec:methods_cc} to facilitate comparison of our upper limit He mass to the maximum total He mass range of $0.06-0.14M_{\odot}$ resulting from the \gls{hach12} model sequence. While the \gls{hach12} model sequence involved modifying the density profile to have a more extended envelope, we obtain He limits within the CO21 model of SN~1994I. For each epoch of SN1994I (we use the same observed data as in \gls{hach12}), we modify the abundance structure of the \gls{hach12} models by replacing the outermost shells with a pure He shell. We perform the He shell replacement in such a way as to conserve the ejecta mass and note that the modification to the abundance structure from the \gls{hach12} models is relatively minor. We repeat this modification for various masses of the outer He shell and each time generate synthetic spectra using \gls{tardis}. We determine an upper limit on the outer He shell mass by comparing the \gls{tardis} synthetic spectra to the observed SN1994I spectra, specifically focusing on the He absorption wavelength regions. The upper He mass limit is chosen such that synthesized TARDIS spectra from models with outer He shell masses larger than that mass show noticeable He features that are not present in the observed spectra of SN1994I.

\section{Results} \label{sec:results}

\subsection{Code Comparison}
\begin{figure}
    \includegraphics[scale=0.35]{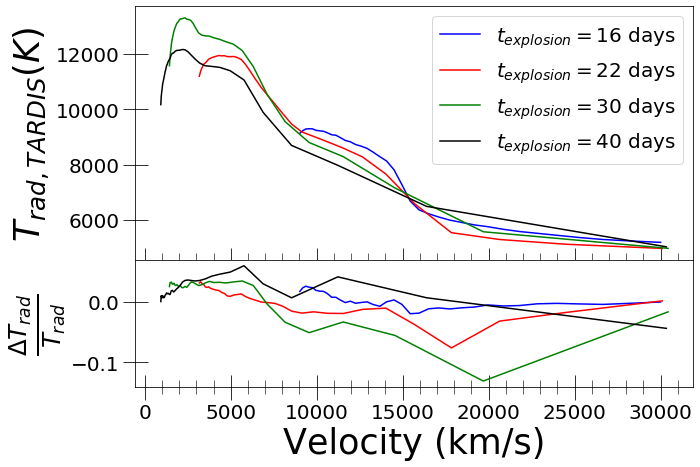}
    \caption{Comparison of the radiative temperature profiles from \gls{tardis} and \gls{hach12} as a function of ejecta velocity. Ejecta shells at higher velocities correspond to larger radii (due to homologous expansion) and encompass in total more mass. We note that $T_{rad}$ is not the same as the temperature of the photosphere. The main panel shows the \gls{tardis} $T_{rad}$ profiles for each of the SN1994I epochs considered in this paper. The lower panel shows the fractional difference $(T_{rad,\gls{tardis}} - T_{rad, Hachinger})/T_{rad,\gls{tardis}}$. The \gls{tardis} converged $T_{rad}$ profile is in strong agreement with that of  \gls{hach12}.}
    \label{fig:trad_prof}
\end{figure}

\begin{figure*}[ht!]
    \includegraphics[scale=0.4]{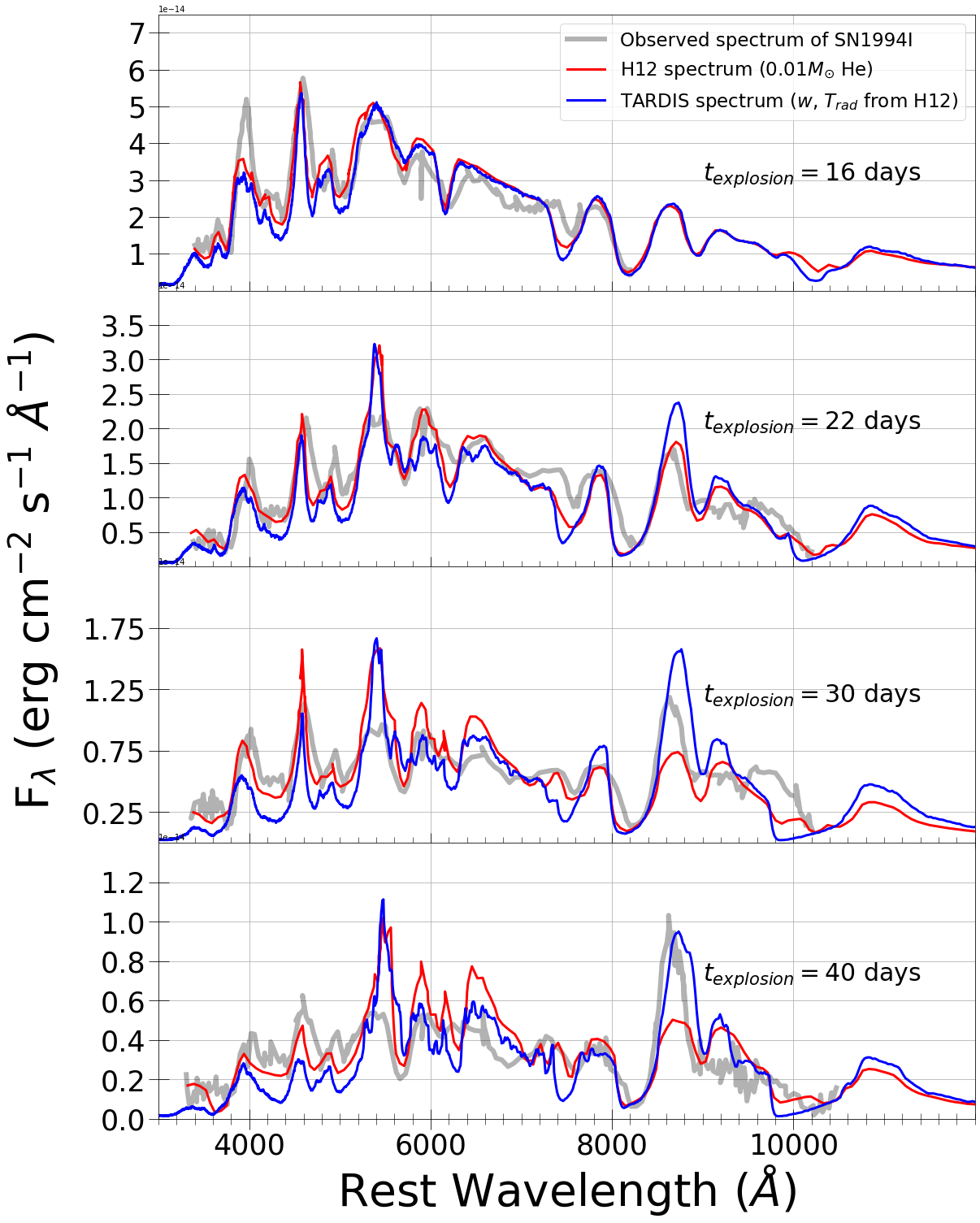}
    \caption{Direct comparison of the \gls{tardis} synthetic spectra (blue) to the \gls{hach12} synthetic spectra (red) for SN1994I. Observed spectra for SN1994I (grey) are provided for reference \citep{filippenko1995type,clocchiatti96}. \gls{tardis} uses the abundance structure, $T_{rad}$ profile, and dilution factor, $w$, profile from \gls{hach12} and the settings from Table \ref{tab:hach_direct_table}. There is generally good agreement between the models.}
    \label{fig:hach_cc_tardis}
\end{figure*}

\label{sec:results_cc}
In order to thoroughly test \gls{tardis} against the code from \gls{hach12}, we compare both the converged radiative temperature profiles ($T_{rad}$), as well as the outputted synthetic spectra. For the $T_{rad}$ comparison, we use the same ejecta models from \gls{hach12} and allow \gls{tardis} to self-consistently solve for a converged plasma state of the SN ejecta. This allows us to test the \gls{tardis} internal plasma state calculations. The synthetic spectra comparison is also run using the same ejecta models from \gls{hach12}, but we also force \gls{tardis} to use the converged $T_{rad}$ and dilution factor ($w$) profiles from \gls{hach12} in order to focus the comparison on the implementation details of MC photon packet propagation. 

Figure \ref{fig:trad_prof} shows the converged \gls{tardis} radiative temperature profiles (main panel) and the fractional difference between the \gls{tardis} and \gls{hach12} temperature profile (lower panel). The radiative temperatures in each shell are the effective temperature of the radiation field exposed to that shell, where the initial blackbody energy distribution emanating from the inner boundary is modified by atomic interactions. At later times, the inner boundary velocity recedes, exposing hotter material closer to the center of the ejecta. We note that $T_{rad}$ is not the same as the temperature of the photosphere. For Figure \ref{fig:trad_prof} we allow \gls{tardis} to self-consistently solve for the radiative temperature of the ejecta shells instead of using the $T_{rad}$ profile obtained from the authors of \gls{hach12}. The fractional difference between the \gls{tardis} and \gls{hach12} converged $T_{rad}$ profiles is on the order of 10 percent, with even better agreement for the early time (16 day and 22 day) synthetic spectra. At later times when the inner boundary approximation to the photosphere is less appropriate, the codes may perform slightly differently due to minor implementation differences, but the fractional difference even for the 30 day and 40 day spectra is still relatively small. Thus we find good general agreement between the \gls{tardis} and \gls{hach12} $T_{rad}$ profile calculations.

While the $T_{rad}$ profile comparison showed general consistency in the radiation field and plasma calculations, we require another comparison to test the MC photon packet propagation between \gls{tardis} and the MC code from \gls{hach12}. For this test we use the \gls{hach12} input models ($\sim 0.01M_{\odot}$ He) and compare the outputted synthetic spectra from both codes which are comprised of MC photon packets that have propagated through the SN ejecta and escaped. We force \gls{tardis} to use the $T_{rad}$ and $w$ profiles from \gls{hach12} in order to focus the comparison on the opacity calculations of each code. We also use the comparison of the synthetic spectra to check for consistency in the He I regions ($\lambda 5876, 6678, 7065, 10831$) in order to validate the NLTE approximation used by \gls{tardis}. Figure \ref{fig:hach_cc_tardis} shows the \gls{tardis} synthetic spectra (blue) compared to the \gls{hach12} synthetic spectra (red) at four epochs. The observed spectra (grey) are provided for reference. There is strong agreement between the codes for the 16 day and 22 day synthetic spectra. In particular, the optical He I regions and the He I 10831 regions are generally consistent between the two synthetic spectra, validating the \gls{tardis} approximation for non-thermal He excitation \citep{boyle_tardis}. Noticeable differences between the synthetic spectra emerge in the later epochs of 30 and 40 days. The locations and widths of the synthetic spectral features generally match, but there are flux offsets, particularly in the Ca H \& K and NIR triplet regions. The smaller flux offsets in the late time synthetic spectra are most likely due to minor differences in atomic data and/or algorithmic implementation exacerbated by the breakdown of the inner boundary approximation to the photosphere for late times. A detailed analysis of the \gls{tardis} photon packets shows that the larger differences in the Ca H \& K and NIR triplet regions is most likely due to fluorescence, but such detailed information for the individual MC photon packets is not available for the \gls{hach12} synthetic spectra. We note that the \gls{tardis} model provides a better description of the observed Ca NIR triplet region (including what appears to be an emission feature) in SN~1994I.

\subsection{Hidden Helium Investigation} \label{sec:results_he}
\begin{figure*}[ht!]
    \includegraphics[width=\textwidth]{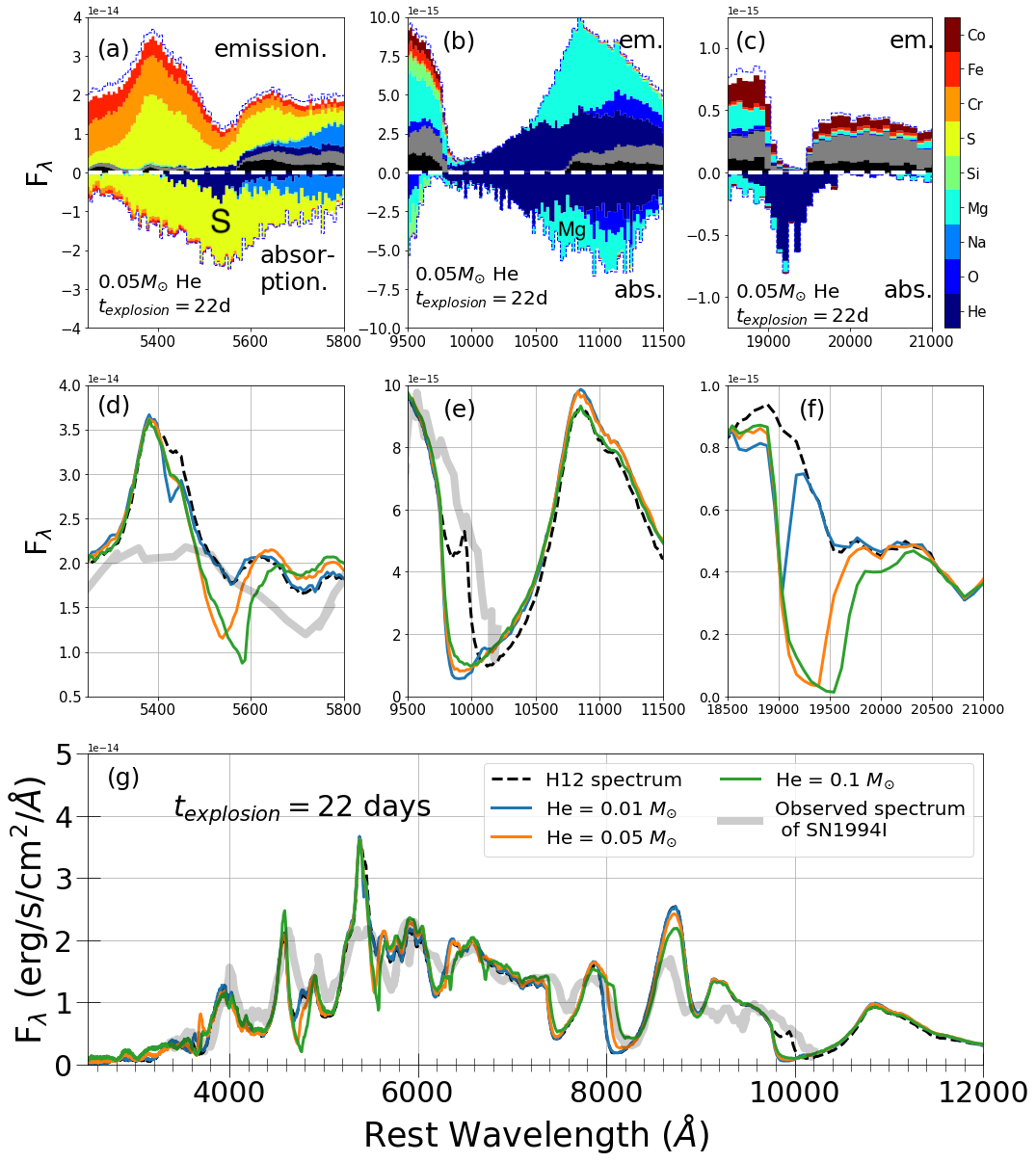}
    \caption{\textit{(a,b,c)}: Photon packet analysis of the SN1994I $t_{explosion}=22$ days TARDIS synthetic spectrum with an outer He shell of $0.05M_{\odot}$. Colors above the dashed white line in the emission (em.) region indicate the element of the last interaction experienced by photons in a given wavelength bin. Black indicates contributions from photons that do not experience any interaction, and grey indicates photons where the only interaction is electron scattering. Colors below the dashed white line in the absorption (abs.) region indicate the last element responsible for removing a photon packet from a given wavelength bin via flourescence. \textit{(d,e,f)}: Zoom-ins of the He I 5876, He I 10831, and He I 20581 features from the synthetic spectra plotted in panel (g). \textit{(g)}: TARDIS synthetic spectra for SN1994I using the H12 abundance model (black, dashed) compared to modified models to include outer He shells of masses $0.01M_{\odot}$ (blue), $0.05M_{\odot}$ (orange), $0.10M_{\odot}$ (green). The observed spectrum is shown for reference (grey). Panel (c) shows that the He I 20581 line is uncontaminated by other elements, and panel (f) shows that the He I 20581 line is extremely sensitive to the outer He shell mass.}
    \label{fig:22d_He_wall}
\end{figure*}

In order to put an upper limit on the amount of hidden He that could exist in the outer layer of the SN1994I ejecta, we systematically insert progressively more massive ($0.001M_{\odot}$, $0.01M_{\odot}$, $0.05M_{\odot}$, $0.1M_{\odot}$) pure He shells into the models from \gls{hach12} to test when noticeable He features appear in the synthetic spectra produced by \gls{tardis}. As we will show, we find that for SN1994I an outer He layer with mass as low as $0.05M_{\odot}$ causes noticeable He features in the synthetic spectra that do not match observations. This is somewhat surprising given the low $^{56}$Ni mixing used by \cite{iwamoto94} which should make it easier to hide He in the outer layer of the SN~1994I ejecta. Figure \ref{fig:22d_He_wall} shows a detailed analysis of the He investigation for the $t_{explosion}=22$ day SN1994I spectrum (we choose this spectrum because it is approximately 2 weeks after maximum light which is when He features are most apparent in observed SNe Ib. See \cite{liu16} for more details). We note that there are trace amounts of He in the inner layers totalling $\sim 0.005M_{\odot}$ which were part of the original \gls{hach12} model, but the contribution of this inner He to the synthetic spectrum is minimal compared to the He present in the outer layer. 

Since we are focusing on the question of hidden He, Figure \ref{fig:22d_He_wall} includes multiple panels zooming in on the He I 5876, 10831, and 20851 regions in the synthetic spectra. In order to securely identify the synthetic spectral features with the elements in the ejecta model that produce them, we plot the last element responsible for interaction with a MC photon packet in the topmost panels (a,b,c). These upper panels show the analysis of which elements contribute to the flux in each wavelength bin. Colors above the dashed white line (drawn at $F_{\lambda}=0$) indicate the element of the last interaction experienced by a photon packet in a given wavelength bin, thus indicating the last atom from which MC photon packets comprising the synthetic spectrum emerged. Colors below the white dashed line indicate the element last responsible for moving a MC photon packet out of a given wavelength bin via flourescence, thus indicating the element last responsible for absorption. If a single element is responsible for the majority of the absorption, then it is reasonable to identify the corresponding synthetic spectral feature as being produced by that element. We note that the average number of interactions an MC photon packet experiences is $\sim 1$, so the last interaction is generally the only interaction (i.e., there are very few packets which experience multiple interactions). Henceforth we refer to the multi-colored plots shown in panels (a,b,c) of Figure \ref{fig:22d_He_wall} as spectral element decomposition plots (first introduced by \cite{kromerplot}, see their Figure 6). In order to determine the mass of the outer He layer for which noticeable He features appear in the synthetic spectra, the middle panels (d,e,f) of Figure \ref{fig:22d_He_wall} show zoom-ins of the He I 5876, 10831, and 20851 regions from each of the synthetic spectra shown in panel (g). We can see that an obvious He I feature that is not present in the observed SN~1994I spectrum appears in panel (d) in the \gls{tardis} spectra with the low-mass $0.05M_{\odot}$ He layer (we discuss the He I 10831 feature in more detail below).

\begin{deluxetable}{CCC}[ht!]
\tablecaption{He I 20581 Pseudo Equivalent Width\label{tab:he_table}}
\tablecolumns{3}
\tablenum{2}
\tablewidth{0pt}
\tablehead{
\colhead{Phase\tablenotemark{a}} &
\colhead{Injected He Mass} &
\colhead{pEW} \\
\colhead{(days)} &
\colhead{$(M_{\odot})$} &
\colhead{$(\AA)$}}
\startdata
16 & 0.001 & N/A\tablenotemark{b}\\
   & 0.01 & 64.03\\
   & 0.05 & 409.25\\
   & 0.10 & 609.52\\
\hline
22 & 0.001 & 145.53\\
   & 0.01 & 62.84\\
   & 0.05 & 482.64\\
   & 0.10 & 729.90\\
\hline
30 & 0.001 & 449.74\\
   & 0.01 & 446.13\\
   & 0.05 & 1182.32\\
   & 0.10 & 1187.99\\
\hline
40 & 0.001 & 555.53\\
   & 0.01 & 500.72\\
   & 0.05 & 453.35\\
   & 0.10 & 1492.34
\enddata

\tablenotetext{a}{Phase measured relative to date of explosion.}
\tablenotetext{b}{There is no identifiable trough in the synthetic spectrum.}

\end{deluxetable}

\begin{figure*}[ht!]
    \includegraphics[width=\textwidth]{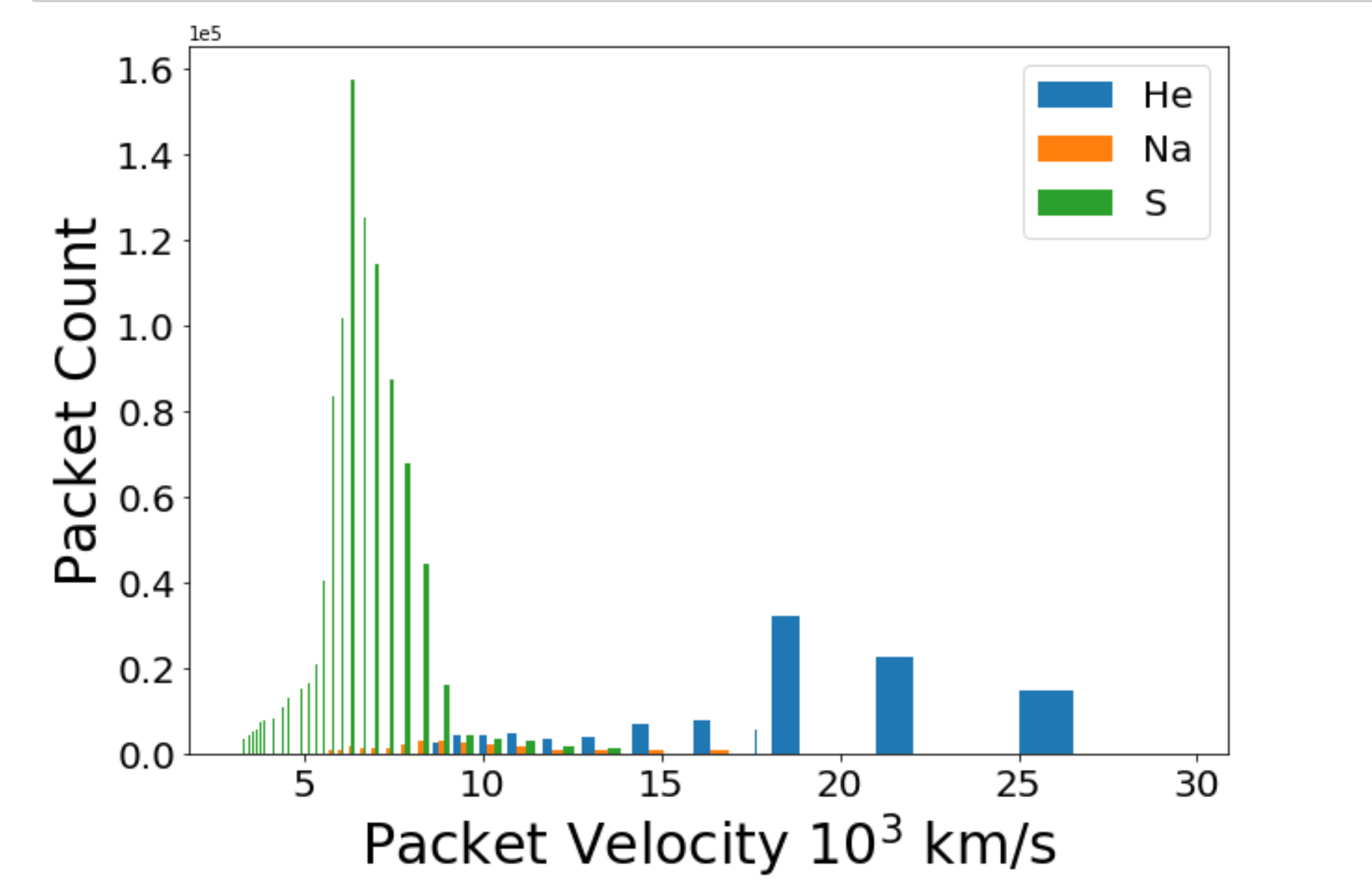}
    \caption{The velocity distributions of where the last interactions occurred for He (blue), Na (orange), and S (green) in the 22 day SN1994I model with an outer He shell of mass $0.05M_{\odot}$. Low velocity (i.e. inner) He is negligible compared to the outer shell He. The distribution of S interactions is much deeper inside the ejecta than the high velocity He interactions.}
    \label{fig:packet_vel}
\end{figure*}

In order to better understand the changes in the synthetic spectra shown in panels (d,e,f) as the mass of the inserted outer He layer increases, we produce the element spectral decomposition plots shown in panels (a,b,c) for each He outer layer mass (only the He mass = $0.05M_{\odot}$ panels are shown in Figure \ref{fig:22d_He_wall}). By examining the photon packets that escape the \gls{tardis} simulations and contribute towards each synthetic spectrum we find that the depth of the trough at $\sim \lambda 5550\AA$ increases as the outer He shell mass increases (see panel (d) of Fig. \ref{fig:22d_He_wall}) due to a larger number of photon packets being absorbed by He. However, it is clear from panel (a) that Sulfur (yellow) is responsible for the majority of the absorption that produces the synthetic spectral feature at $\sim \lambda 5550\AA$. In particular, examining the \gls{tardis} MC photon packets contributing to this feature reveals that the S II 5640 and S II 5606 lines dominate the absorption from He I 5876 and Na I D for this model of SN~1994I. While the blending of the Na I D and He I 5876 lines is well documented \citep{patat01, dessart12}, particularly with regard to SN1994I \citep{clocchiatti96,sauer94Imodel}, Sulphur is not discussed in the literature as a potential contaminant for the possible He I 5876 line in SNe Ic. This is somewhat surprising as it has been shown that S is produced through oxygen burning in shock-driven nucleosynthesis for core-collapse supernovae \citep{woosley1995evolution,woosley1995presupernova}, and S is present in realistic models of SNe Ic \citep{dessart15}. In order for the S II 5640 and 5606 lines to contaminate the He I 5876 and Na I D lines (which are usually produced at high velocities in the SN ejecta), the S must be located at lower velocities in the ejecta. Figure \ref{fig:packet_vel} shows the velocity distributions of where in the SN~1994I ejecta the last bound-bound interactions occurred for He, Na, and S. As expected (since the S II 5604 and 5606 lines are bluer than He I 5876 and Na I D), we find that S line interactions occur much deeper in the ejecta than He and Na. This further complicates the He I 5876 line identification question in SNe Ic because the strengthening over time of the weak absorption feature seen in some observed spectra could be due to increased non-thermal excitation of He, blending with Na, or blending with lower velocity S as the photosphere recedes. Thus, we give supporting evidence to earlier suggestions that the SN Ib classification should not be based on just the He I 5876 line \citep{liu16} given the contamination by other lines.

Just as we used the synthetic spectral element decomposition plots to analyze the potential He I 5876 feature in our \gls{tardis} spectra, we apply a similar analysis to the potential NIR He I 10831 absorption feature. The identification of an absorption feature in the spectra of SN~1994I with He I 10831 has been thoroughly debated. There is some evidence that the inferred velocity of the potential He I 10831 feature is consistent with the velocities of extremely weak optical He I lines \citep{filippenko1995type,clocchiatti96}, but modelers have managed to reproduce the observed NIR absorption feature using Si I \citep{millard99} and a mixture of He I and C I \citep{Baron94Imodel,sauer94Imodel}. In this work we show that high velocity Mg can also be a contaminant of the He I 10831 feature, confirming a hypothesis from \cite{filippenko1995type} and work by \cite{dessart15}. In particular, panel (b) of Figure \ref{fig:22d_He_wall} shows that the He I 10831 region is contaminated with Mg (cyan), confirming that it is not a good line for inferring the presence or mass of He in SN~1994I.

The reaction of the He I 10831 absorption feature in panel (e) of Figure \ref{fig:22d_He_wall} in response to increasing the mass of the inserted outer He layer appears counter-intuitive because the potential He I 10831 trough weakens as the outer He layer gains mass. However, a detailed analysis of the MC photon packets reveals that the number of packets absorbed by He actually increases with increasing outer He shell mass. The trough in the synthetic spectrum weakens instead of strengthening because by inserting the pure He shell, we are replacing some outer Mg which was contributing to the absorption comprising this feature. Thus the net effect is that the trough in the 1 micron region becomes slightly shallower with increasing He outer shell mass.

Finally, we apply the synthetic spectral element decomposition analysis shown in Panel (c) of Figure \ref{fig:22d_He_wall} for day 22 of SN~1994I with a $0.05M_{\odot}$ outer He shell to each of the \gls{tardis} simulations shown in Panel (f). The He I 20851 line is considered the most unambiguous indicator of He due to the lack of other nearby lines to act as contaminants \citep{dessart15}. Our \gls{tardis} simulations for SN~1994I confirm that the He I 20851 region is uncontaminated, as shown in Panel (c) of Figure \ref{fig:22d_He_wall}. The colors below the dashed white line show that He is the dominant contribution towards the 2$\mu$m trough. Specifically, more than 97 percent of the absorption between 1.9$\mu$m and 1.95$\mu$m is due to He. There are no observed spectra of SN1994I including this feature, but our \gls{tardis} simulations predict that the He I 20581 feature can be used for inferring outer He shell mass for future observations of SNe Ic. Panel (f) of Figure \ref{fig:22d_He_wall} shows that the He I 20851 line consistently changes as outer He shell mass increases. In table \ref{tab:he_table} we calculate the pseudo equivalent width (pEW) of the He I 20581 absorption feature in our synthetic spectra using the method from \cite{liu16}, and we show that there is a clear trend of pEW increasing with increasing outer He shell mass. This result demonstrates the importance of Near Infrared (NIR) spectroscopy for constraining the He abundance of SNe Ic. To our knowledge, this is the first time that modelling has indicated that the outer He shell mass could be inferred from observations of the He I 20851 feature for SN~1994I-like objects.

\section{Summary}\label{sec:summary}
In this paper, we present the first application of \gls{tardis} to a SN Ic, 1994I. We perform a thorough investigation to determine an upper bound of $0.05M_{\odot}$ for the mass of an outer He shell where obvious optical and NIR He features appear in the spectra. This result indicates that SN~1994I is almost completely He deficient in comparison with typical He masses (0.5-1.0$M_{\odot}$, \cite{dessart15}) of He-rich SNe progenitors, although no observed spectra cover the He I 20851 line which would provide the most confident assessment of He abundance. We present evidence for the first time that low velocity S II lines can be major contributors to blending of the He I 5876 line in SNe Ic. Moreover, the strengthening of He I 5876 absorption due to non-thermal effects is degenerate with the strengthening of the S II lines due to the recession of the photosphere over time. Finally, we show that not only is the He I 20851 line uncontaminated by other elements, but that it can also be used to infer the outer He shell mass using the pEW of the absorption feature. This highlights the critical importance of obtaining NIR spectra of SNe Ic in order to address the hidden helium question for a statistical sample of SNe. Future work will involve dedicated \gls{tardis} modeling of the recent SN Ic 2020oi where NIR spectra are available \citep{rho_2020oi}. The SN~2020oi modeling will use the recently developed \gls{tardis} emulator \citep{kerzendorf_dalek} which will enable us to produce posterior distributions for the He abundance. In addition, we will work to extend the \gls{tardis} modeling capabilities to include new types of explosive phenomena with similar radiative transfer needs like fast blue optical transients \citep{modjaz19_review, pritchard_18gep}.

\section{Acknowledgements}
The authors are extremely grateful to Stephan Hachinger for discussions on the differences between the \gls{tardis} synthetic spectra and his work. We are also grateful for Dr. Hachinger's willing collaboration in sending us the initial conditions used in his simulations. M.M. and the SNYU group are supported by the NSF CAREER award AST-1352405, by the NSF award AST-1413260 and by a Humboldt Faculty Fellowship. We thank the Google Summer of Code program for supporting student contributions to \gls{tardis}.

\bibliography{williamson.bib}
\end{document}